\documentclass[12pt,preprint]{aastex}

\begin{document}

\title{The 5 GHz Arecibo Search for Radio Flares from Ultracool Dwarfs }

\author{Matthew Route\altaffilmark{1,2} \& Alexander Wolszczan\altaffilmark{1,2}}

\altaffiltext{1}{Department of Astronomy and Astrophysics, the Pennsylvania State University, 525 Davey Laboratory, University Park, PA 16802, mroute@astro.psu.edu, alex@astro.psu.edu}

\altaffiltext{2}{Center for Exoplanets and Habitable Worlds, the Pennsylvania State University, 525 Davey Laboratory, University Park, PA 16802}

\slugcomment{Accepted for publication in ApJ}

\begin{abstract}
We present the results of a 4.75 GHz survey of 33 brown dwarfs and one young exoplanetary system for flaring radio emission, conducted with the 305-m Arecibo radio telescope.  The goal of this program was to detect and characterize the magnetic fields of objects cooler than spectral type L3.5, the coolest brown dwarf detected prior to our survey.  We have also attempted to detect flaring radio emission from the HR 8799 planetary system, guided by theoretical work indicating that hot, massive exoplanets may have strong magnetic fields capable of generating radio emission at GHz frequencies. We have detected and confirmed radio flares from the T6.5 dwarf 2MASS J10475385+2124234. This detection dramatically extends the temperature range over which brown dwarfs appear to be at least sporadic radio-emitters, from ~1900 K (L3.5) down to ~900 K (T6.5).  It also demonstrates that the utility of radio detection as a unique tool to study the magnetic fields of substellar objects extends to the coolest dwarfs, and, plausibly to hot, massive exoplanets. We have also identified a single, 3.6$\sigma$ flare from the L1 dwarf, 2MASS J1439284+192915. This detection is tentative and requires confirmation by additional monitoring observations.
\end{abstract}

 
\section{Introduction}

Radio detection of low-mass stars and brown dwarfs of the spectral types of M7 and beyond, commonly called ultracool dwarfs (UCDs), provides a unique observational method to investigate the magnetic field generation, and its structure and stability in these objects \citep{ber09,doy10}. The combination of low temperatures and fast rotation of the UCDs makes other observing methods, such as the detection of X-ray and H$\alpha$ activity and spectropolarimatry inefficient or simply impossible to use \citep{mcl12,mor10,reb10}.

Despite the meager detection rate of about 9\%, of both the quiescent and the flaring radio emission from the M7-T8.5 UCDs observed prior to this survey \citep{ant13}, the existing data have been sufficient to reveal a clear dependence of the observed radio activity on stellar rotation rate \citep{ber08b,mcl12}. Observations of radio flares from some UCDs that are most likely generated by the electron cyclotron maser instability (ECMI, Treumann 2006) indicate that these rapid rotators maintain kilo-Gauss magnetic fields \citep{hal07,hal08}. In fact the recent detection of flares from the T6.5-dwarf, 2MASS J10475385+2124234 (hereafter J1047+21), (Route \& Wolszczan 2012, this paper) suggests that the UCDs populating the LT spectral range also maintain fields of that strength. In addition, the still rare series of measurements of individual UCDs that extend over several years (e.g. TVLM 513-46546, Doyle et al. 2010) indicate that these fields are stable over comparable periods of time or, very likely, much longer. Although much theoretical work has been done to model the magnetic field generation in fully convective UCDs, its detailed physics remains enigmatic \citep{dur93,cha06,dob06}. The same is true for the physics and geometry of ECMI generation in these sources \citep{nic12,yue12}.

On general grounds, as argued by \citet{rec10}, the strength and evolution of magnetic fields in rapidly rotating stars, brown dwarfs, and exoplanets can be modeled in terms of a scaling law that governs the available energy flux as a function of mass, luminosity and radius. One intriguing consequence of this model is that young, super-Jupiter-mass exoplanets could have magnetic fields bordering the 1 kGauss range, which could make such planets detectable at frequencies around and above 1 GHz. Observations of radio emission from young exoplanets, in conjunction with similar measurements made over the entire UCD spectral range, would provide a unique experimental basis for studies of the dynamo generation and magnetic field properties of bodies across the transition from low-mass stars to brown dwarfs, over the LTY spectral range, and through a much more ``fuzzy'' border between brown dwarfs and planets \citep{hal13}.

The Arecibo UCD survey described in this paper has been directly motivated by the absence of radio detections of these objects beyond the L3.5 spectral type, and by the need to test, if at least some of the LT-dwarfs that have temperatures similar to those of hot, young exoplanets, are detectable radio emitters. A good example of a potentially observable planetary system is provided by the HR 8799 planets, which have temperatures in the 800 K - 1100 K range, and their central A-star is not expected to be radio active \citep{gra99,mar08}. Such detections would provide the much needed evidence that the magnetic properties of both the coolest brown dwarfs and the hot exoplanets could be studied using their radio emission as a diagnostic tool. This would provide a way to constrain the models of dynamo generation and hence internal structure of these objects. In addition, placing useful constraints on the existence and properties of magnetic fields of exoplanets would obviously enrich the studies of planetary habitability \citep{sca07}.

We have used the superb sensitivity of the 305-m Arecibo telescope to rapidly varying radio signals, and the broadband capability of its 5 GHz receiving system, to conduct a search for flaring radio emission from a sample of the LT-dwarfs in the instrument's declination range, and from the HR 8799 planetary system. In this paper, we discuss the initial phase of this survey and its results. We describe the target list and the observing method in Section 2. The results are presented in Section 3 and further discussed in Section 4.

\section{Target Selection and Observations}

In a typical general survey of ultracool dwarfs \citep{ant08,ant13,bur05,ber06,mcl12}, the quiescent component of radio emission from an UCD is first detected as a point source in a deep synthesis map, and then the time series of measured flux densities is searched for any variability. However, Arecibo is a single, fixed-dish telescope, and has a restricted practical sensitivity to weak, quiescent emission from radio sources. This is due to a large beam width and the related confusion limitations, the presence of man-made interference, and difficulties in obtaining a precise calibration of ON-OFF-source measurements caused by the telescope aperture blockage and signal reflections from the surrounding terrain. On the other hand, because of the large collecting area, broad receiver bandwidth, and the manner in which Stokes parameters are calculated from the orthogonally polarized outputs \citep{hei02}, the telescope is an excellent instrument for detection of broadband, circularly polarized radio emission that varies both in time and frequency. In this case, the calibration by position switching is not necessary and the signal is almost interference-free, except for occasional strong bursts that are easy to identify.

Given these characteristics of the telescope, our recently completed LT-type UCD survey discussed here was aimed at the detection of significantly ($>$ 10\%) polarized, rapidly varying (flaring) emission from these objects. We have chosen 33 targets from the list of UCDs that have had no previous detections of quiescent or flaring radio emission, their declinations fell in the 0$^\circ$ - 38$^\circ$ Arecibo range, and that were less than 40 pc away of the Sun, as given in Table 1. Three of these targets, 2MASSI J0700366+315726, 2MASS J09373487+2931409, and 2MASS J07271824+1710012 were previously observed by \citet{ant13} without any detection.  As the observed flares from some UCDs are many times brighter than their quiescent emission (e.g. DENIS 1048-3956 \citep{bur05}, J1047+21 \citep{rou12}), and the Arecibo telescope is highly sensitive to this type of radio emission, our choice of a wider than typical distance range was reasonable. For example, the 30 mJy flare from DENIS 1048-3956, and relatively frequent $\sim$20 mJy flares from TVLM513-46546 \citep{hal09}, if emitted from 40 pc away, would still represent $\geq$3$\sigma$ detections with this survey's sensitivity. In the target list presented in Table 1, all but four objects have spectral types $\ge$L3.5, down to T8.5, which is the spectral range that had no detections of the quiescent and/or flaring radio emission at the commencement of the survey. We have also added the two well known flaring objects, TVLM 513-46546 (hereafter TVLM 513) \citep{ber02,ber08a,for09,doy10,hal06,hal07,ost06} and 2MASSW J0746425+200032 (hereafter J0746+20) \citep{ant08,ber09}, for testing and calibration purposes. Finally, for reasons discussed earlier, our list included the young HR 8799 planetary system \citep{mar08,mar10}.

Observations were made between January 2010 and April 2012 with the 305-m Arecibo radio telescope equipped with the 5 GHz, dual-linear polarization receiving system. For measurements at low zenith angles, the antenna gain and the system temperature were $\sim$8 K/Jy and $\sim$30 K, respectively. During each observation, we recorded the time resolved (dynamic) spectra at the target position for about 1 to 2 hours of the available tracking time of the telescope. Typically, such observations were repeated several times for each target to accumulate the total on-source times listed in Table 2. The spectra for the four Stokes parameters were sampled at 0.1 s intervals by seven 172 MHz-bandwidth, 8192-channel, FPGA-based Mock spectrometers \citep{sal09}, tuned to cover a $\sim$1 GHz bandpass around the center frequency of 4.75 GHz. 

\begin{figure}
\centering
\includegraphics[width=0.8\textwidth,angle=0]{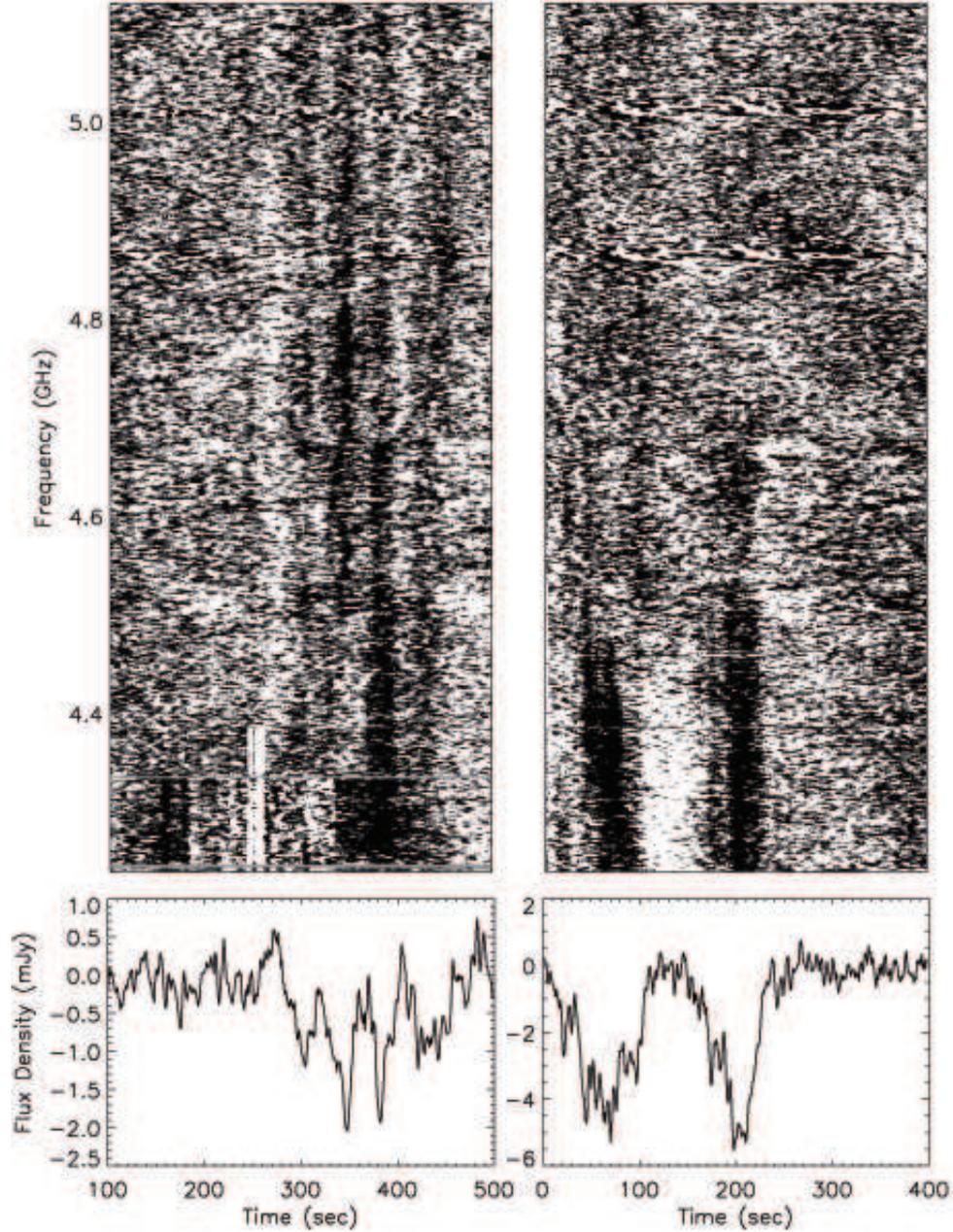}
\caption{Dynamic spectra (top) and average time profiles (bottom) of the circularly polarized flares from the UCDs TVLM-513 (left) and J0746+20 (right) smoothed to the time and frequency resolution of 6.3 s and 2.1 MHz, respectively.  Both spectra display the characteristic time - frequency structure observed in stellar flares (e.g. Osten \& Bastian 2008), and also previously seen in TVLM-513 \citep{hal09}. The pair of white lines in the TVLM-513 spectrum near t=240 s is left over from the RFI excision.  \label{fig1}}
\end{figure}

Data analysis involved formation of the dynamic spectra of the four interference-cleaned and calibrated Stokes outputs generated by the spectrometers. These spectra were then inspected at different time and frequency resolutions in search of rapid flux density variations. Typically, we used the time and frequency resolution ranges of 0.9 s to 30 s and 2 MHz to 172 MHz, respectively. The dynamic spectra and bandpass-averaged, Stokes V burst profiles of our calibrator targets, TVLM 513 and J0746+20, are shown in Fig. \ref{fig1} as an example. They display the characteristic bands of frequency drifting emission, also seen in J1047+21 \citep{rou12}. These features were used in data analysis as a criterion to distinguish genuine UCD flares from radio interference.

The nominal, 1$\sigma$ sensitivity of our observations to broadband bursts integrated over all the seven, 172 MHz spectrometer sub-bands, and binned to 0.9 s resolution, was $\sim$0.15 mJy.  However, the band-limited nature of the flares, and a persistent presence of radio interference, especially in the lowest and the highest frequency sub-bands, restricted the practical 1$\sigma$ detection sensitivity of the survey to $\sim$0.4 - 1.2 mJy. In fact, the 0.9 s - 172 MHz time-frequency resolution combination, typically used in the initial data inspection, represents a good empirical tradeoff between the high sensitivity requirement and the ability to detect the bright, short timescale structures often present in the bursts (Fig. 1).

\section{Results}

Our survey has detected two new radio flaring UCDs: the T6.5 brown dwarf, J1047+21 (Fig. 2a), and the L1 dwarf, 2MASS J1439284+192915 (J1439+19 hereafter, Fig. 2b). An extensive discussion of flares from J1047+21, the coolest radio emitting dwarf detected so far, has been given by \citet{rou12} and will not be repeated here. However, we note that, guided by our observations, \citet{wil13} have detected quasi-quiescent, apparently unpolarized, 16.5 $\mu$Jy emission from J1047+21 with the Jansky Very Large Array (JVLA) at 6 GHz. This implies that deeper imaging may lead to detections of many UCDs, for which only upper limits to a quiescent emission have been derived so far (e.g. McLean et al. 2012, Antonova et al. 2013, and references therein).

J1439+19 was initially detected in the Two Micron All-Sky Survey (2MASS) archive, and was selected for follow up study on the basis of its red J-Ks color \citep{kir99}. It was classified by these authors as an L1 brown dwarf due to the nearly equivalent strength of its TiO, CrH, and FeH absorption features. Trigonometric parallax gives its distance as 14.37 pc, and it has an effective temperature of 2273 K \citep{dah02}. Kinematic arguments suggest that its age is approximately 3.2 Gyr \citep{sei10}. J1439+19 appears to be single, as no companion has been detected out to $\sim$4.3 AU and $>$0.2 mass ratio \citep{rei08}. It has a rotational velocity, $v$sin$i$, of 11.1 km s$^{-1}$ \citep{bla10}. Previous radio observations of the source conducted with the VLA suggested that this ultracool dwarf was radio-quiet, with a 78 $\mu$Jy upper limit to its radio flux \citep{mcl12}.

The weak, $\sim$1 mJy burst from J1439+19 is 90\% circularly polarized, and it peaks up around 4.3 GHz, fading away toward the higher frequencies. Because the presence of the signal in the dynamic spectrum is marginal, we consider this detection as a tentative one, still requiring confirmation by followup observations. If real, the detected burst could have been generated by the electron cyclotron maser instability (ECMI) \citep{tre06}, as it appears to be the case of bursts from other UCDs \citep{ant08,ber09,hal06,hal07,hal08,mcl11}. Under this assumption, the magnetic field strength, $B$, of J1439+19 can be estimated from $\nu_{c} = 2.8 \times 10^{6}$ B (Gauss), where $\nu_{c}$, the local electron cyclotron frequency, is taken to be equal to 4.3 $\times$ 10$^{9}$ Hz, the center frequency of the flare. Given the frequency extent of the observed emission, one obtains $B\geq 1.5$ kG, which is similar to the values derived for other ultracool dwarfs.

The observed and derived parameters of the two detected sources, J1047+21 and J1439+19, and the upper detection limits for the remaining objects are listed in Table 2, and the dynamic spectra of the bursts are depicted in Figure 2. We have taken a conservative approach to the computation of these limits, and defined them as 3$\sigma$ rms noise limits derived from the signal that was smoothed to the 0.9-s resolution and integrated over the cleanest out of the seven available spectrometer sub-bands. As discussed in Section 2, this solution was dictated by both the expected time-frequency behavior of the flares (Figs. 1,2), and the presence of radio interference.

\section{Discussion}

The detection of highly polarized flares from the T6.5 dwarf, J1047+21, by this survey \citep{rou12} provides the first example of an UCD that remains capable of the generation of bursts of radio emission despite its largely neutral atmosphere, as dictated by the star's very low, $\sim$900 K temperature. Further evidence that J1047+21 maintains an ionized environment needed for this kind of activity comes from the fact that it is both a weak H$\alpha$ emitter \citep{bur03}, and the source of a very low-level, quasi-quiescent radio emission \citep{wil13}.

Of course, it is possible that the observed radio flares from J1047+21 represent an exception and that flares from the UCDs in the LT spectral range are extremely rare. However, the \citet{wil13} result suggests that pushing the detection limits below the $\sim$10-20 $\mu$Jy limit may lead to new discoveries. Further encouragement comes from the recent detection of a quiescent, 0.4 mJy emission at 5.5 GHz from the L5+T7 binary, 2MASS J13153094-2649513AB, by \citet{bur13}.

Our 3.6$\sigma$ detection of a highly circularly polarized, broadband, short duration signal from the direction of J1439+19 appears significant. The characteristics of this $\sim$1 mJy burst do not match those of any known radio interference signals in the Arecibo 5 GHz band. It is visible in both Stokes I and V and has the overall appearance that makes it similar to flares from other UCDs detected by this survey and the previous ones. However, in absence of additional detections, the authenticity of this signal is obviously questionable. Confirmation of radio flares from J1439+19 is important, because it would make this object a flaring UCD with the lowest value of $v sini$ (Table 1). Only two other quiescent radio sources, J0952210-192431 \citep{mcl12} and J14563983-280947 \citep{bur05}, have significantly lower values of this parameter at 6 km s$^{-1}$ \citep{rei02} and 8 km s$^{-1}$ \citep{reb10}, respectively. As all the known flaring UCDs are rapid rotators \citep{mcl12}, such a confirmed detection could help improve our understanding of the role of rotation in UCD magnetic field generation.

Our 3.5-hour observation of the HR 8799 planetary system has not produced any obvious detection. It was motivated by the surprising radio activity of J1047+21 described above, and by the theoretical possibility that young, massive, rapidly rotating ``super-Jupiter'' planets could have magnetic fields that would be strong enough to generate radio bursts in the GHz range \citep{rec10}. Prospects to detect radio emission from this system with the JVLA have been also considered by \citet{ost11} but, to our knowledge, no such detection has been accomplished so far. Nevertheless, with the temperatures of the four substellar bodies orbiting HR 8799 in the 800 K - 1100 K range, masses from 5 M$_{Jup}$ to over 30 M$_{Jup}$, depending on the assumed age of the system, and the T-dwarf or even Saturn-like spectra \citep{opp13}, this system remains a very exciting target for future radio observations over a possibly wide frequency range.

Surveys of a total of 224 UCDs of the M7-T spectral types have detected 15 sources of radio emission, predominantly in the 5-8 GHz frequency range (Antonova et al. 2008, 2013; Berger 2002, 2006; Burgasser \& Putman 2005; Burgasser et al. 2013; McLean et al. 2011, 2012; Phan-Bao 2007; this paper). With one exception (the Australia Telescope Compact Array (ATCA) search by \citet{bur05}), all the published surveys have been carried out with the VLA and all of them used broadly similar criteria of target selection, the same observing frequencies ($\sim$5 GHz and $\sim$8 GHz) and the identical observing method (deep aperture synthesis mapping). Because our survey had no practical sensitivity to quiescent emission, we can only compare it to the previous ones that examined the time series for all targets, regardless of the initial detection or non-detection of them as quiescent radio sources \citep{ant08,ant13,ber02,ber06,bur05,mcl12}. By this accounting, the total number of UCDs, whose data have been analyzed in this manner reduces to 96, and, out of that number, nine objects have been found to be variable radio emitters. In addition, five of these have been confirmed to vary periodically, in synchronism with stellar rotation, at periods ranging from 2 to 4 hours. They include the flaring UCDs, TVLM 513 \citep{ber02,ber08a,for09,doy10,hal06,hal07,ost06}, LSR J1835+3259 \citep{hal08}, 2MASS J0746+20 \citep{ant08,ber09}, and the two other objects, 2MASS J00361617+1821104 \citep{hal08}, and 2MASS J13142039+1320011 \citep{mcl11}, both of which show slow, periodic intensity variations that do not have a rapid, low duty cycle flaring character. 

If we exclude the latter two UCDs, because our survey was not sensitive to a quiescent or slowly varying emission, the approximate detection probability of the rapidly flaring sources is $\sim$7\%. Given the large uncertainty involved in this estimate, it is in a rough agreement with the detection rate of the published surveys that have been sensitive to low duty cycle flaring, including the $\sim$6\% efficiency of our program.

Plausible astrophysical reasons for rarity of both the quiescent and the flaring radio emission from UCDs have been recently summarized by \citet{ant13}. These include geometric effects due to beaming in the case of outbursts powered by the ECMI mechanism, low duty cycle of the outbursts, details of the UCD magnetic field and radio emission generation, and its possible long-term variability. In addition, for the UCDs that flare in synchronism with stellar rotation, the ability to detect such flares obviously depends on the rotation phase coverage by observations. Not surprisingly, all authors of the published UCD surveys agree that longer observations of more objects over a wider frequency range are needed to gain a better understanding of the physics behind the observed phenomena. In this context, it is important to reiterate that not all the time series data accumulated by the previous UCD surveys have been routinely analyzed with a sufficient resolution to maximize the sensitivity to a rapidly varying and/or flaring emission. If a flare of a duration $\Delta$t occurs within an observing window extending over time $T$, and $\Delta$t$\ll$T, then the resultant signal-to-noise ratio (S/N) of the mean signal will be reduced by a factor of ($\Delta$t/T)$^{1/2}$. As an example, the S/N of a $\sim$5 min. flare, observed as a continuum source over the 1-3 hour integration time, typically used in the VLA UCD mapping observations, would be reduced by a factor of $\sim$3-6.

An improved sensitivity to flaring in future observations may be particularly important because of the possibility that UCDs of the coolest spectral types may be predominantly sporadic (sometimes periodic), rather than quiescent radio emitters. Although this contradicts the currently available evidence that almost all flaring UCDs are also quiescent radio sources, it may represent a bias introduced by the prevailing manner of data analysis, in which a presence of flaring is checked for in the time series data only if an UCD has been detected as a quiescent emitter. In addition, as demonstrated by the radio properties of J1047+21 \citep{rou12,wil13}, the coolest dwarfs may be easier to detect while flaring, because of the extremely low levels of quiescent emission.
 
\section{Acknowledgements}

We thank Phil Perillat for his assistance with observations and data reduction. MR acknowledges support from the Center for Exoplanets and Habitable Worlds and the Zaccheus Daniel Fellowship. The Center for Exoplanets and Habitable Worlds is supported by the Pennsylvania State University, and the Eberly College of Science. The Arecibo Observatory is operated by SRI International under a cooperative agreement with the National Science Foundation (AST-1100968), and in alliance with Ana G. M\'{e}ndez-Universidad Metropolitana, and the Universities Space Research Association.  


\clearpage

\begin{deluxetable}{llllllll}
\rotate
\tabletypesize{\footnotesize}
\tablecolumns{8}
\tablewidth{0pt}
\tablecaption{Survey Target Properties}
\tablehead{
	\colhead{Name}&
	\colhead{R.A.}&
	\colhead{Dec.}&
	\colhead{Spectral Type}&
	\colhead{Distance}&
	\colhead{v $sin i$}&
	\colhead{Radio Flux}&
	\colhead{Reference}\\
	\colhead{}&
	\colhead{({hh}\phn{mm}\phn{ss})}&
	\colhead{(\phn{\arcdeg}~\phn{\arcmin}~\phn{\arcsec})}&
	\colhead{}&
	\colhead{(pc)}&
	\colhead{(km s$^{-1}$)}&
	\colhead{($\mu$Jy)}&
	\colhead{}
}
\startdata
J00325937+1410371 & 00 32 59 & +14 10 37 & L8 & 33.18 &  &  & 1\\
J0039191+211516 & 00 39 19 & +21 15 17 & T7.5 & 11.11 &  &  & 2\\
J01514155+1244300 & 01 51 41 & +12 44 30 & T1 & 21.4 &  &  & 1\\
J02074284+0000564 & 02 07 42 & +00 00 56 & T4.5 & 28.69 &  &  & 1,3\\
J0326137+295015 & 03 26 14 & +29 50 15 & L3.5 & 32.26 &  &  & 4\\
J03284265+2302051 & 03 28 42 & +23 02 05 & L8 & 30.18 &  &  & 5\\
J03454316+2540233  & 03 45 43 & +25 40 23 & L0 & 26.95 &  &  & 6\\
J07003664+3157266  & 07 00 37 & +31 57 27 & L3.5 & 12.2 & 29.9 & $<$78 & 7, 8\\
J07271824+1710012  & 07 27 18 & +17 10 01 & T8 & 9.08 &  &  & 9\\
J08251958+2115521  & 08 25 20 & +21 15 52 & L7.5 & 10.66 &  & $<$45 & 5\\
J0850359+105716  & 08 50 36 & +10 57 16 & L6 & 25.58 &  &  & 4\\
J085911+101017  & 08 59 11 & +10 10 17 & T7 & 26-36 &  &  & 10\\
J09121469+1459396  & 09 12 14 & +14 59 39 & L8 & 20.48 &  &  & 11,12\\
J09373487+2931409  & 09 37 34 & +29 31 40 & T7 & 6.14 &  &  & 9, 13\\
J10475385+2124234  & 10 47 53 & +21 24 23 & T6.5 & 10.56 &  & $<$45 & 3, 14\\
J11122567+3548131  & 11 12 26 & +35 48 13 & L4.5 & 21.72 & 28.7 &  & 5, 8, 12\\
J11463449+223053  & 11 46 35 & +22 30 53 & L3 & 27.17 & 23.9 &  & 4, 15, 16\\
J115739+092201  & 11 57 39 & +09 22 01 & T2.5 & 24-34 &  &  & 10\\
J123828+095351  & 12 38 29 & +09 53 51 & T8.5 & 20 &  &  & 17, 18\\
J131508+082627  & 13 15 08 & +08 26 27 & T7.5 & 19-28 &  &  & 10\\
J1328550+211449  & 13 28 55 & +21 14 49 & L5 & 32.26 &  &  & 4\\
J133553+113005  & 13 35 53 & +11 30 05 & T9 & 10-10.6 &  &  & 18\\
J14243909+0917104  & 14 24 39 & +09 17 10 & L4 & 31.55 & 32.5 & $<$96 & 19, 20, 21\\
J1439284+192915  & 14 39 28 & +19 29 15 & L1 & 14.37 & 11.1 &  & 4, 8, 15\\
J144601+0024519  & 14 46 01 & +00 24 52 & L6 & 22 &  &  & 1, 3\\
J15232263+3014562  & 15 23 22 & +30 14 56 & L8 & 18.62 &  & $<$45 & 5, 12\\
J16241436+0029158  & 16 24 14 & +00 29 15 & T6 & 11 & 36.6 & $<$36 & 22, 23, 24\\
J16322911+1904407  & 16 32 29 & +19 04 40 & L8 & 15.24 & 21.8 & $<$54 & 4, 15, 21, 25\\
J1711457+223204  & 17 11 46 & +22 32 04 & L6.5 & 30.2 &  &  & 3, 5\\
J1841086+311727  & 18 41 09 & +31 17 27 & L4 & 40 &  &  & 3, 5\\
J21011544+1756586  & 21 01 15 & +17 56 58 & L7.5 & 33.18 &  &  & 3, 5\\
HR 8799 b,c,d,e  & 23 07 29 & +21 08 03 &  & 39.4 &  &  & 26\\
\enddata
\tablecomments{{\bf References.} (1) \citet{geb02}; (2) \citet{mug06}; (3) \citet{vrb04}; (4) \citet{kir99}; (5) \citet{kir00}; (6) \citet{kir97}; (7) \citet{tho03}; (8) \citet{bla10}; (9) \citet{bur02}; (10) \citet{pin08}; (11) \citet{wil01}; (12) \citet{per97}; (13) \citet{sch09}; (14) \citet{bur99}; (15) \citet{dah02}; (16) \citet{bai01}; (17) \citet{kir12}; (18) \citet{bur08}; (19) \citet{fah09}; (20) \citet{bec88}; (21) \citet{moh03}; (22) \citet{str99}; (23) \citet{tin03}; (24) \citet{zapo06}; (25) \citet{bas00}; (26) \citet{bow10}.  This table benefited from ``The M, L, T, and Y dwarf compendium,'' DwarfArchives.org, 6 Dec. 2002.}
\end{deluxetable}
\clearpage

\begin{deluxetable}{llllllll}
\rotate
\tabletypesize{\footnotesize}
\tablecolumns{7}
\tablewidth{0pt}
\tablecaption{Survey Detection Results}
\tablehead{
	\colhead{Object}&
	\colhead{Spectral Type}&
	\colhead{Time on Source}&
	\colhead{Detected Flux}&
	\colhead{L$_{rad}$}&
	\colhead{L$_{bol}$\,\tablenotemark{a}}&
	\colhead{L$_{rad}$/L$_{bol}$}\\
	\colhead{}&
	\colhead{}&
	\colhead{(ks)}&
	\colhead{(mJy)}&
	\colhead{(log L$_{\odot}$)}&
	\colhead{(log L$_{\odot}$)}&
	\colhead{(log L$_{\odot}$)}
}
\startdata
J00325937+1410371  & L8 & 3.6 & $<$1.101 & -8.773 & -4.55{*} & -4.223\\
J0036159+182110  & L3.5 & 6.6 & $<$1.179 & -9.900 & -4.51 & -5.390\\
J0039191+211516  & T7.5 & 11.4 & $<$1.14 & -9.708 & -5.41{*} & -4.298\\
J01514155+1244300  & T1 & 13.2 & $<$1.262 & -9.094 & -4.54{*} & -4.554\\
J02074284+0000564  & T4.5 & 4.2 & $<$3.135 & -8.445 & -4.74{*} & -3.705\\
J0326137+295015  & L3.5 & 7.8 & $<$1.293 & -8.727 & -4.62{*} & -4.107\\
J03284265+2302051  & L8 & 14.4 & $<$1.044 & -8.878 & -4.55{*} & -4.328\\
J03454316+2540233  & L0 & 9 & $<$1.206 & -8.914 & -3.56 & -5.354\\
J07003664+3157266  & L3.5 & 12 & $<$1.455 & -9.521 & -3.88{*} & -5.641\\
J07271824+1710012  & T8 & 13.8 & $<$1.101 & -9.898 & -5.58{*} & -4.318\\
J07464256+2000321  & L0+L1.5 & 36.6 & 5.1 & -8.949 & -3.93 & -5.019\\
J08251958+2115521  & L7.5 & 7.8 & $<$1.242 & -9.707 & -5.21 & -4.497\\
J0850359+105716  & L6 & 7.2 & $<$1.302 & -8.926 & -4.34{*} & -4.586\\
J085911+101017  & T7 & 7.2 & $<$1.29 & -8.904 & -5.26{*} & -3.644\\
J09121469+1459396  & L8 & 3.6 & $<$1.473 & -9.065 & -4.55{*} & -4.515\\
J09373487+2931409  & T7 & 9 & $<$1.227 & -10.191 & -5.26{*} & -4.931\\
J10475385+2124234  & T6.5 & 103.8 & 2.7 & -9.351 & -5.35 & -4.001\\
J11122567+3548131  & L4.5 & 4.2 & $<$1.473 & -9.014 & -4.16{*} & -4.854\\
J11463449+223053  & L3 & 7.2 & $<$1.146 & -8.929 & -3.96{*} & -4.969\\
J115739+092201  & T2.5 & 7.2 & $<$1.674 & -8.708 & -4.55{*} & -4.158\\
J1238285+095351  & T8.5 & 3.6 & $<$1.026 & -9.243 & $<$-5.58{*} & -3.663\\
J131508+082627  & T7.5 & 7.2 & $<$1.464 & -8.952 & -5.41{*} & -3.542\\
J1328550+211449  & L5 & 7.2 & $<$1.158 & -8.775 & -4.22{*} & -4.555\\
J133553+113005  & T9 & 4.2 & $<$1.242 & -9.737 & -5.58{*} & -4.157\\
J14243909+0917104  & L4 & 7.2 & $<$1.617 & -8.650 & -4.04 & -4.610\\
J1439284+192915  & L1 & 4.2 & 1.062 & -9.489 & -3.67{*} & -5.819\\
J144601+0024519  & L6 & 3.6 & $<$1.098 & -9.131 & -4.34{*} & -4.791\\
J15232263+3014562  & L8 & 5.4 & $<$1.211 & -9.233 & -5.27 & -3.963\\
J16241436+0029158  & T6 & 2.4 & $<$3.474 & -9.233 & -5.16 & -4.073\\
J16322911+1904407  & L8 & 20.4 & $<$1.217 & -9.405 & -5.31 & -4.095\\
J1711457+223204  & L6.5 & 15 & $<$1.109 & -8.851 & -4.39{*} & -4.461\\
J1841086+311727  & L4 & 3 & $<$3.696 & -8.033 & -4.09{*} & -3.943\\
J21011544+1756586  & L7.5 & 11.4 & $<$3.172 & -8.011 & -4.5{*} & -3.511\\
TVLM 513  & M8.5 & 12.6 & 3.07 & -9.142 & -3.59 & -5.552\\
HR 8799  &  & 39.4 & $<$1.044 & -8.647 & -5.1 & -3.547\\
\enddata
\tablenotetext{a}{Asterisks denote bolometric luminosities inferred from \citet{vrb04}. HR 8799 data from \citet{mar08,mar10}. Detected objects TVLM 513 and J0746+20 are calibration sources.}
\end{deluxetable}
\clearpage
\newpage

\begin{figure}
\centering
\includegraphics[width=0.8\textwidth,angle=0]{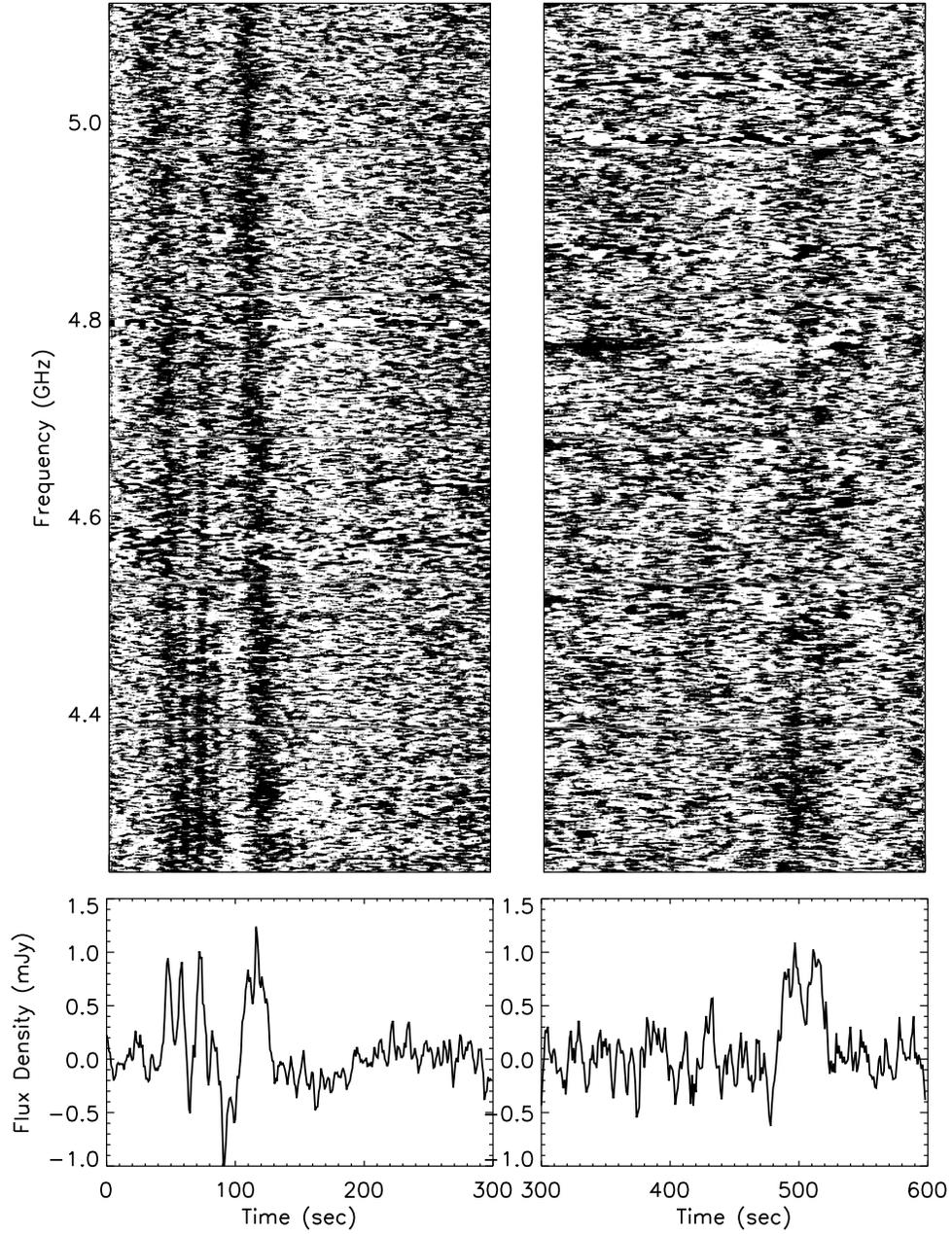}
\caption{Dynamic spectra (top) and average time profiles (bottom) of the circularly polarized flares from the newly detected sources, J1047+21 (left, see also Route \& Wolszczan 2012) and J1439+19 (right). The J1439+19 flare weakens toward higher frequencies, but it remains visible over the entire bandpass. See Fig. 1 for details. \label{fig2}}
\end{figure}

\begin{figure}
\centering
\includegraphics[width=0.8\textwidth,angle=0]{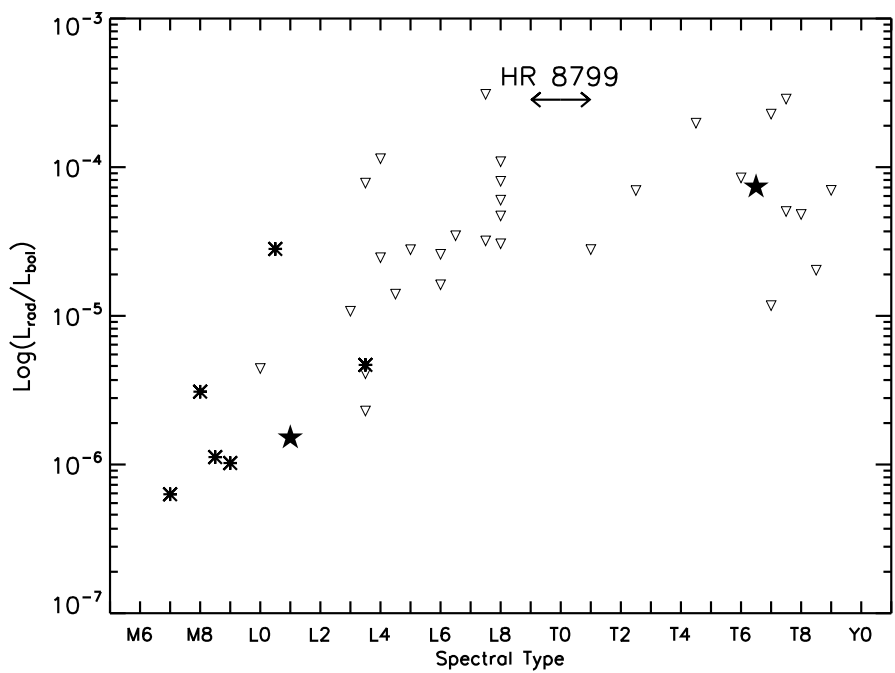}
\caption{A logarithmic plot of the ratio of the radio luminosity to the bolometric luminosity versus spectral type for the UCDs surveyed by this program. Filled stars mark the two detected objects, the L1 dwarf, J1439+19, and the previously reported T6.5 UCD, J1047+21 \citep{rou12}. Inverted open triangles represent the upper limits for the survey targets from Table 2. The horizontal arrows mark the hypothetical location range of the HR 8799 planets, given their measured temperatures and spectral characteristics \citep{opp13}, and the upper limit to their radio luminosities from this paper. Asterisks denote the previous detections of flaring UCDs discussed in the text.\label{fig3}}
\end{figure}

\end{document}